\documentclass[a4paper]{VisionStyle}
\usepackage{epsfig,psbox}

\begin{document}

\title{Properties of the background of EPIC onboard XMM-Newton}

\author{H. Katayama\inst{1}, I. Takahashi\inst{2}, Y. Ikebe\inst{3},
K. Matsushita\inst{3}, Y. Tanaka\inst{3}, and M. Freyberg\inst{3}}

\institute{Department of Earth and Space Science, Graduate School of
Science, Osaka University, 1-1 Machikaneyama, Toyonaka, 560-0043 Osaka,
Japan
\and
Department of Physics, University of Tokyo, 7-3-1 Hongo, Bunkyo-ku,
113-0033 Tokyo, Japan
\and
Max Planck Institut f\"ur Extraterrestrische Physik, Postfach 1312, 85741
Garching, Germany
}
\maketitle

\begin{abstract}

Understanding a background is crucial in particular for a study of low
surface brightness objects. In order to establish the background
subtraction method, we have studied properties of the EPIC background.

Count rates of the background vary violently by two order of magnitude at
the maximum, while during the most quiet period, these are stable within
8 \% at a 1 $\sigma$ level. The overall spectrum is dominated by
particle events above 5 keV, and its spatial variation is also found.

The long-term variation of the background is also investigated with CAL
CLOSED data, which is the data of calibration source with filter closed. 
The average background count rate decreased by 20 \% from March 2000 to
January 2001, but it regained in February 2001.

For the modeling of the background spectrum, we investigate correlations
between the 2-10 keV count rate and some characteristic parameters. The
PN background shows a good correlation with some parameters. On the
other hands, the MOS background does not shows a clear
correlation. Further investigation is needed for the MOS background.

Our final goal is to establish a method to predict the background, for
which these results will be reflected in the background generator.

\end{abstract}

\section{Data and Screening}

In order to study properties of the background,
we used data from Science PV phase data operated in the full frame
mode with thin filter. All the data are shown in table
\ref{tbl:hkatayama-WA3_tbl1} and table \ref{tbl:hkatayama-WA3_tbl2}. The
number of data sets is 8 for PN and 7 for MOS.

\begin{figure}[htbp]
\begin{center}
 \psbox[xsize=0.2#1,ysize=0.2#1]{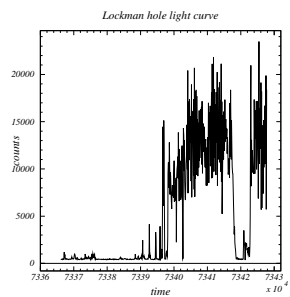}
 \caption{Light curve of Lockman hole observation. }
 \label{fig:hkatayama-WA3_fig1}
\end{center}
\end{figure}

Figure \ref{fig:hkatayama-WA3_fig1} shows a light curve of PN 
from one of the Lockman hole observations. 
In order to exclude the flare events, time periods where the count rate 
deviates from the mean value during
quiescent periods by $\pm2 \sigma$ are excluded. Celestial sources and
noisy columns are also excluded from the data.
The average exposure time thus remained after excluding the flare events
is 18 ks for PN and 26 ks for MOS.

\begin{table}
\begin{center}
  \begin{tabular}{l l l}\hline
  Obs ID	                & Exposure &   Object   \\ \hline\hline
  0063\_0123100201\_PNS001	& 11411	  & MS0737.9+7441 \\
  0070\_0123700101\_PNS003	& 27695	  & Lockman Hole \\ 
  0071\_0123700201\_PNS003	& 21009	  & Lockman Hole \\ 
  0073\_0123700401\_PNS003	& 9977	  & Lockman Hole \\ 
  0078\_0124100101\_PNS003	& 21403	  & RXJ0720.4-3125 \\
  0081\_0123701001\_PNS003	& 17569	  & Lockman Hole \\ 
  0082\_0124900101\_PNS003	& 17665	  & MS1229.2+6430 \\ 
  0181\_0098810101\_PNS003	& 18529	  & WW Hor \\ \hline
  \end{tabular}
  \caption{PN data summary}
  \label{tbl:hkatayama-WA3_tbl1}
\end{center} 
\begin{center}
  \begin{tabular}{l l l}\hline				
  Obs ID	 &              Exposure	& Object \\ \hline\hline
  0070\_0123700101\_M1S001	& 30552	  & Lockman Hole \\ 
  0071\_0123700201\_M1S001	& 33269	  & Lockman Hole \\ 
  0078\_0124100101\_M1S001	& 28794	  & RXJ0720.4-3125 \\ 
  0081\_0123701001\_M1S001	& 21358	  & Lockman Hole \\ 
  0100\_0123701001\_M1S001	& 21258	  & Lockman Hole \\ 
  0071\_0123700201\_M2S002	& 26465	  & Lockman Hole \\ 
  0181\_0098810101\_M2S002	& 21785	  & WW Hor \\ \hline
  \end{tabular}
  \caption{MOS data summary}
  \label{tbl:hkatayama-WA3_tbl2}
\end{center}
\end{table}

\section{Spectrum of background}

\begin{figure}[htbp]
\begin{center}
  \psbox[r,xsize=0.25#1,ysize=0.25#1]{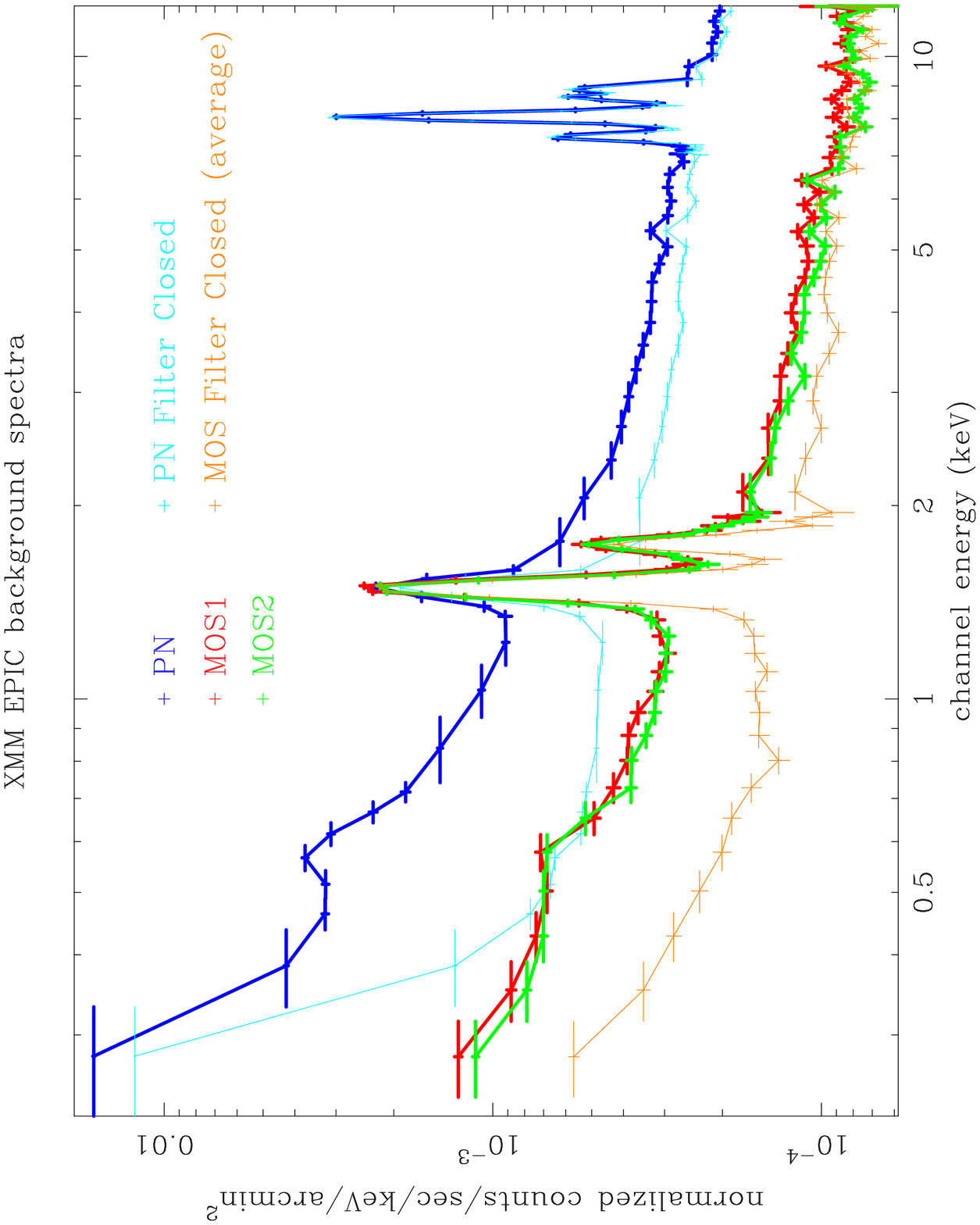}
\caption{Average background spectra of XMM EPIC. Internal backgrounds (Filter
 closed) are also displayed. }
\label{fig:hkatayama-WA3_fig2}
\vspace*{0.5cm}
 \begin{minipage}[cbt]{3.5cm} 
 \psbox[r,xsize=0.17#1,ysize=0.17#1]{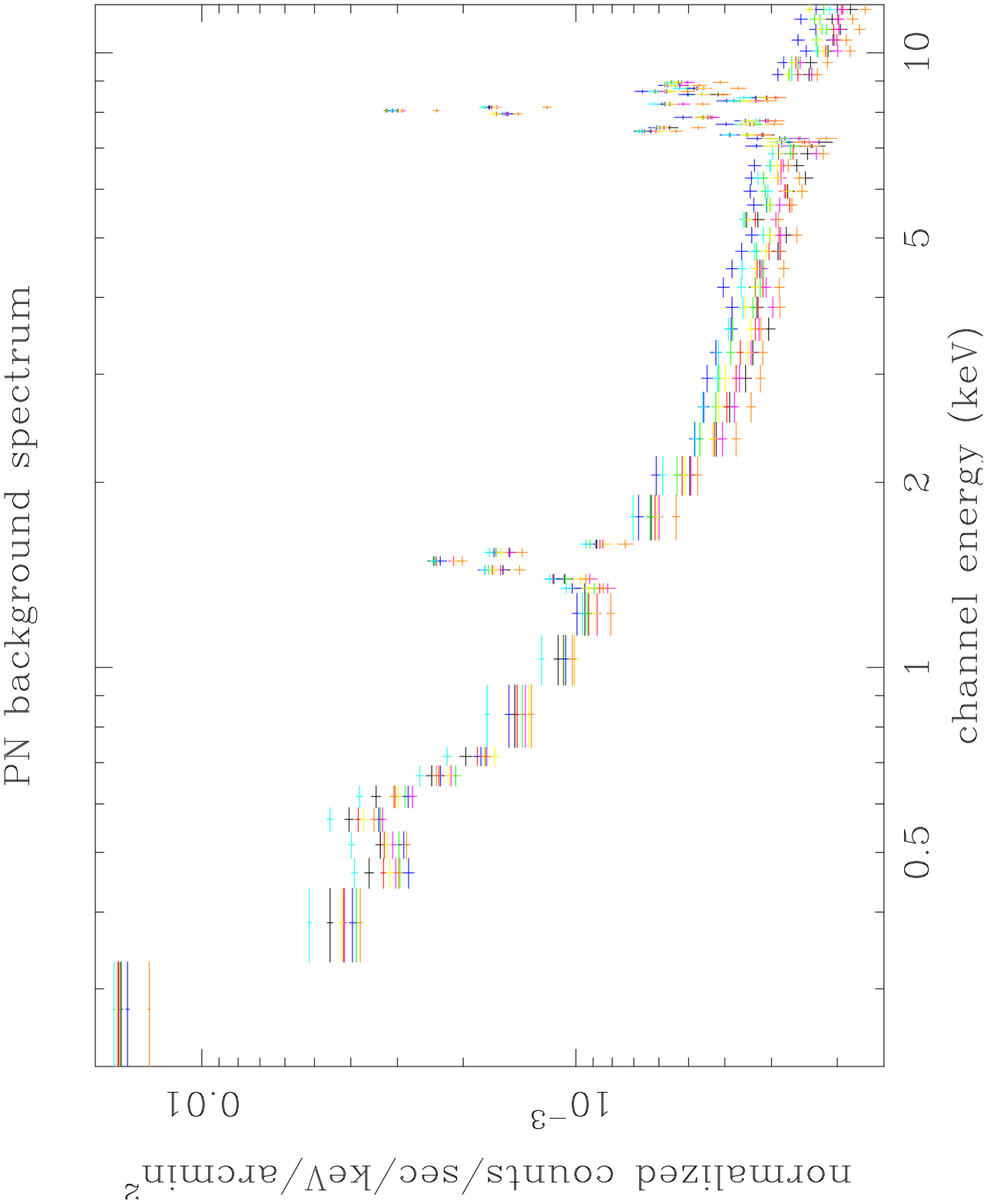}
\vspace*{0.1cm}
 \psbox[r,xsize=0.17#1,ysize=0.17#1]{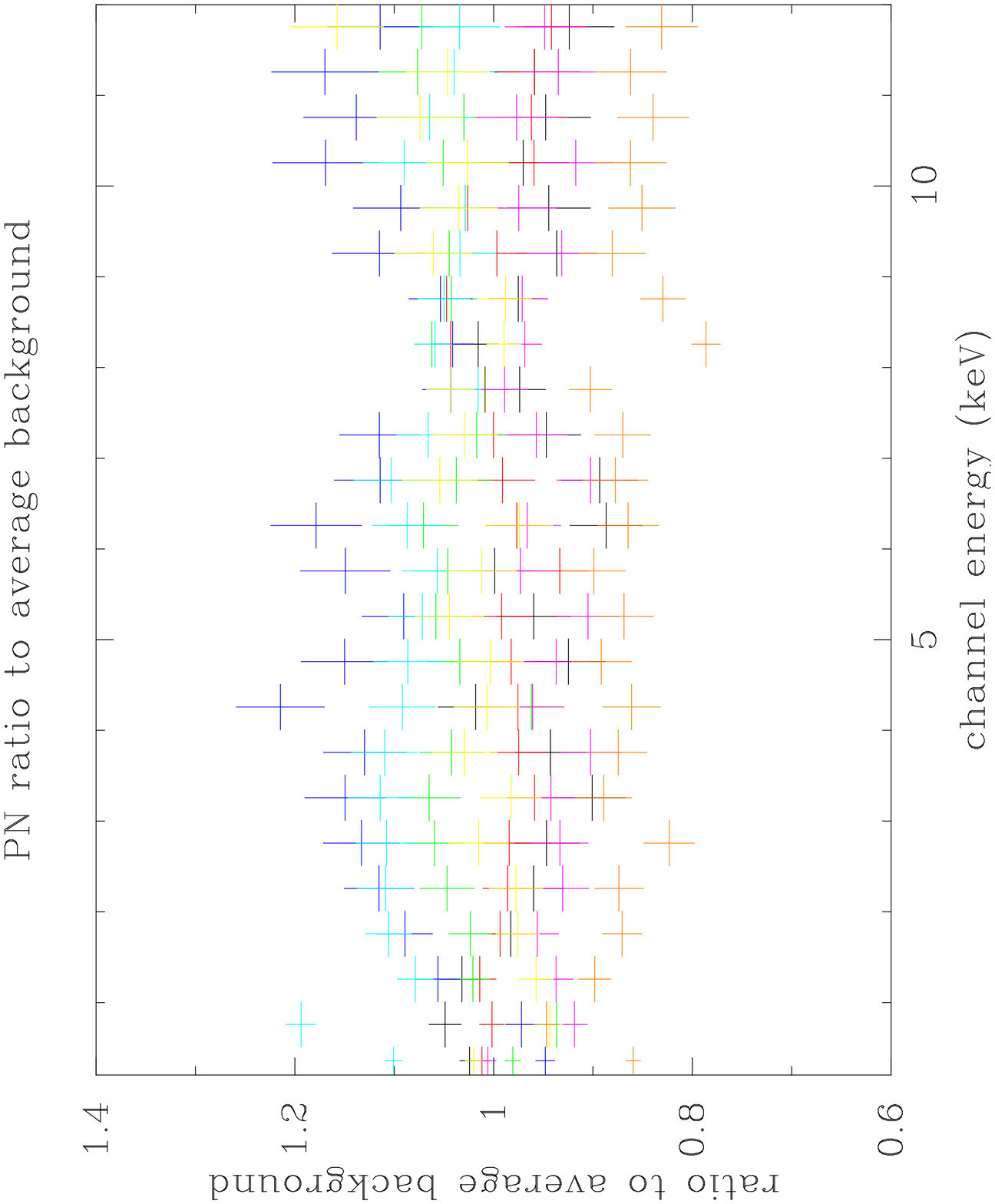}
 \end{minipage}
 \hspace*{0.5cm}
 \begin{minipage}[cbt]{3.5cm} 
 \psbox[r,xsize=0.17#1,ysize=0.17#1]{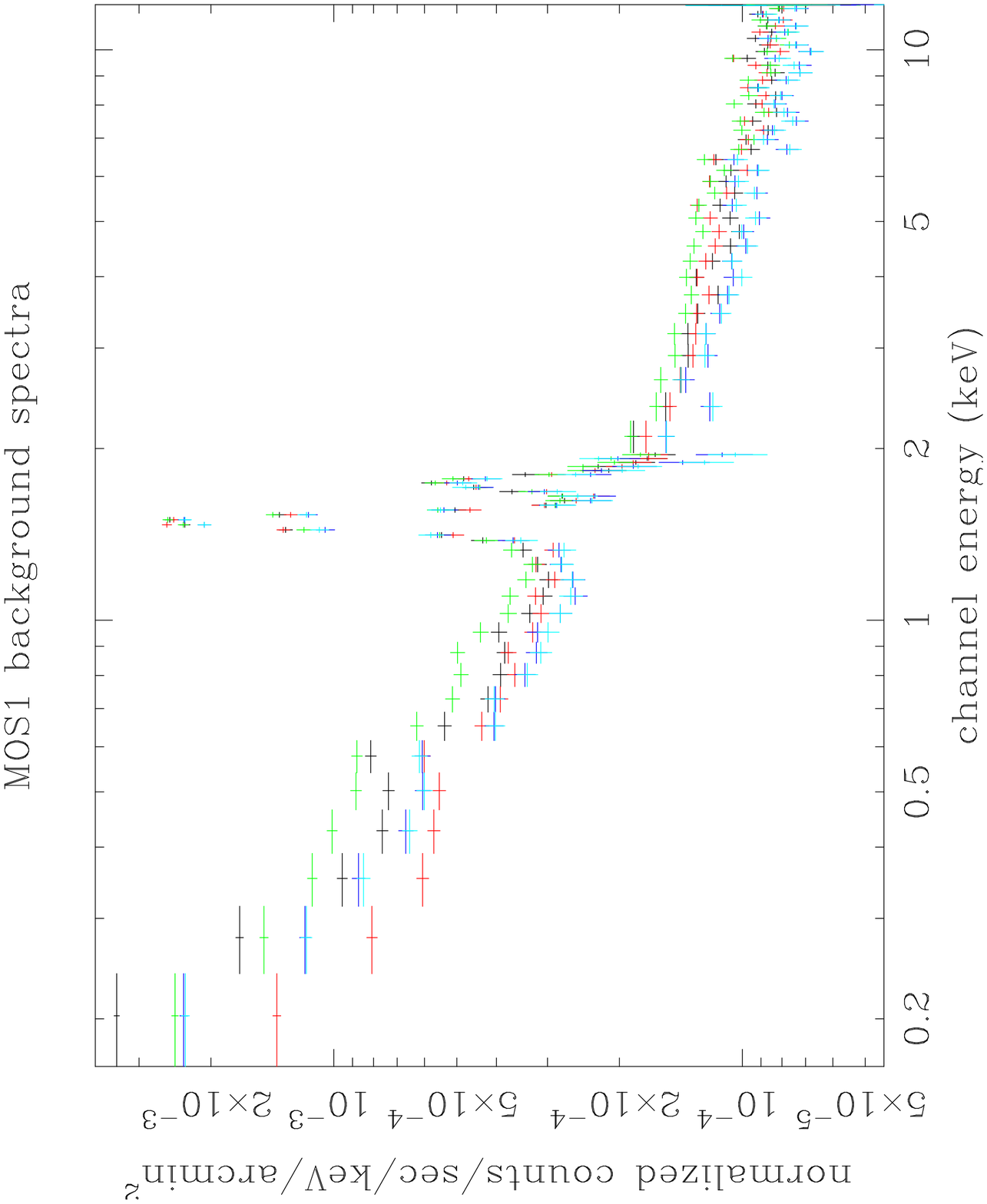}
\vspace*{0.1cm}
 \psbox[r,xsize=0.17#1,ysize=0.17#1]{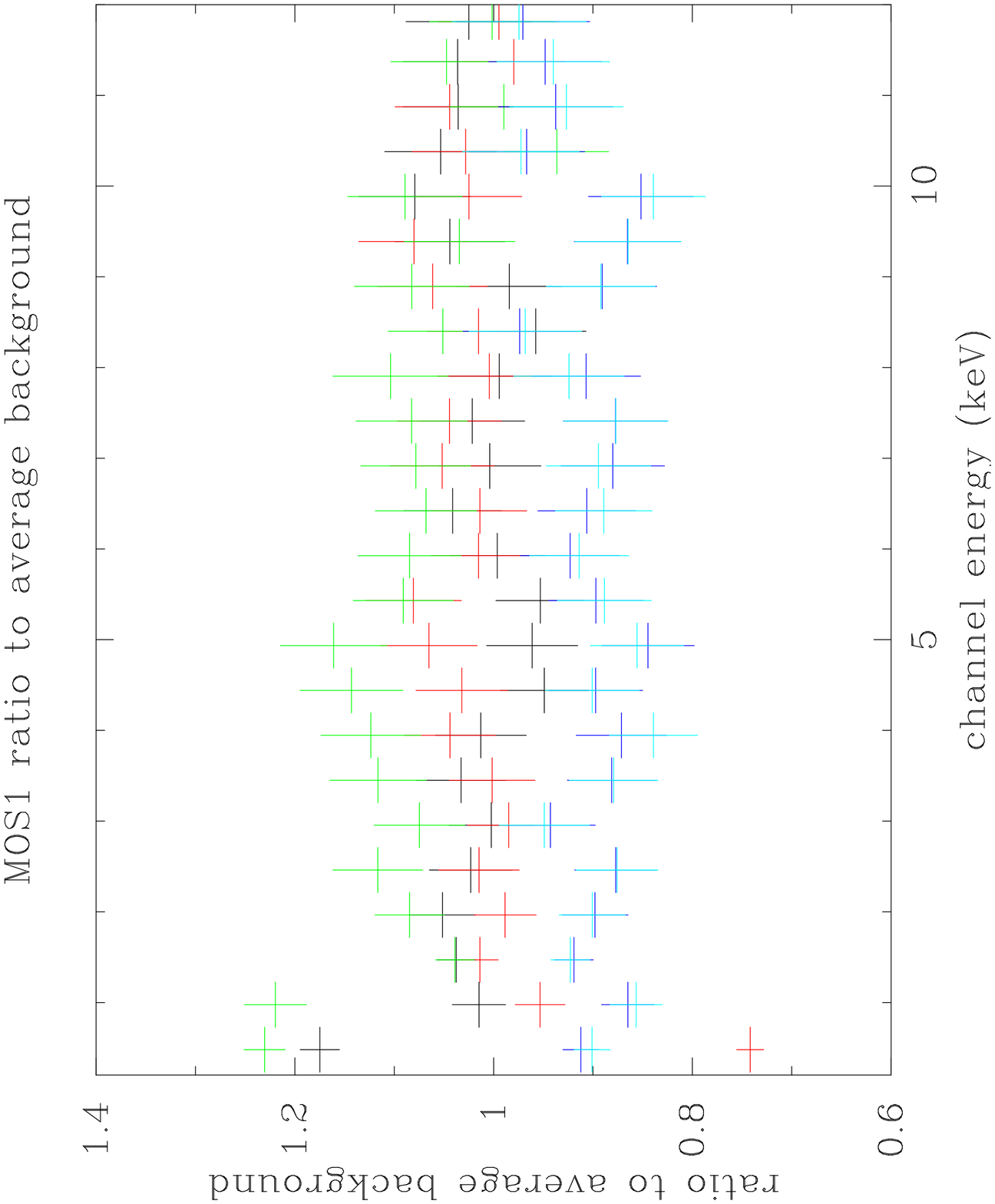}
 \end{minipage}
\end{center} 
\caption{(Top) All background spectra of individual
 observations. Right is PN background spectra and Left is MOS1. (Bottom)
 Ratio to the average background spectrum. Colors are corresponding to
 individual observation of the top figure.}  
\label{fig:hkatayama-WA3_fig3}
\end{figure}

Figure \ref{fig:hkatayama-WA3_fig2} shows average background spectra of
XMM EPIC.  Internal backgrounds which are taken with the filter wheel in
the closed position are also displayed in the same figure. The prominent
features of the PN background spectrum are Al-K, Ni-K, Cu-K, and Zn-K
lines. Two line features of the MOS background spectrum are Al-K and Si-K
lines.  The overall spectrum is dominated by particle
events above 5 keV.

Background spectra from individual observations are compared in
figure \ref{fig:hkatayama-WA3_fig3} (Top). Figure
\ref{fig:hkatayama-WA3_fig3} (Bottom) shows their ratio to the average
spectrum. The fluctuation of spectra is within 8 \% at a 1 $\sigma$ level.

\section{Spatial distribution of background}

\subsection{Average spatial distribution}

\begin{figure}[htbp]
\begin{center}
 \psbox[xsize=0.2#1,ysize=0.2#1]{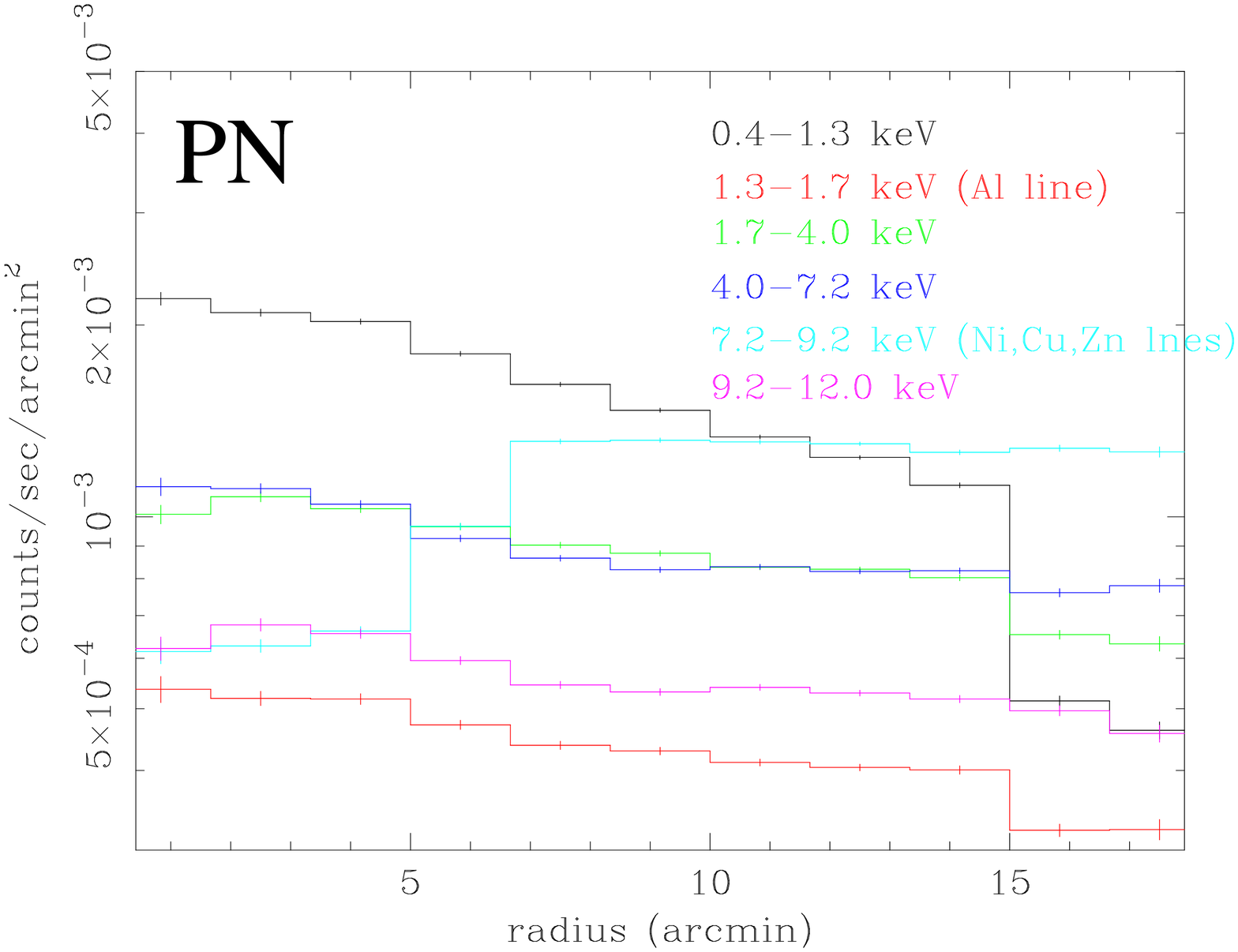}
\vspace*{0.2cm}
 \psbox[xsize=0.2#1,ysize=0.2#1]{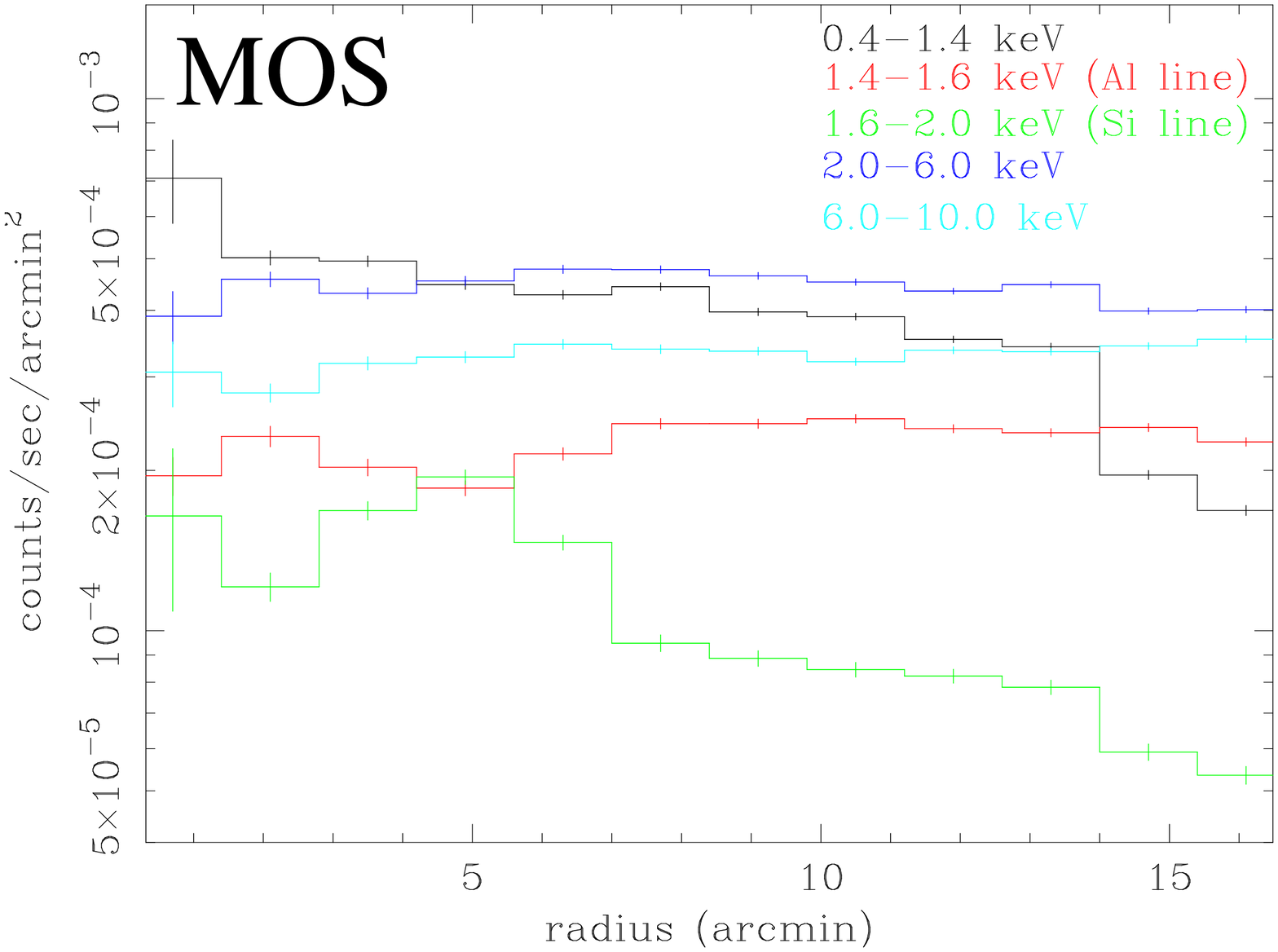}
 \caption{(Top) Average spatial distribution of the PN
 background. (Bottom) Average spatial distribution of the MOS1
 background.}
 \label{fig:hkatayama-WA3_fig4}
\end{center}
\end{figure}

Figure \ref{fig:hkatayama-WA3_fig4} displays the average spatial
distribution of the background in different energy bands. The size of
the FOV for PN and MOS is 15' and 14' radius, respectively.  In the
outside of the FOV, no sky X-ray is supposed to be detected, and the gap
in the count rate profile is prominent in soft energy bands.  Vignetting
effect of telescope is also seen in the lower energy band, while the
continuum emission in the higher energy band makes almost flat spatial
distribution.  Fluorescent lines show peculiar profiles (see XMM User's
hand book and \cite*{hkatayama-WA3:frey01} in this contribution).

\subsection{Correlations of count rate}

\begin{figure}[htbp]
\begin{center}
 \psbox[xsize=0.17#1,ysize=0.17#1]{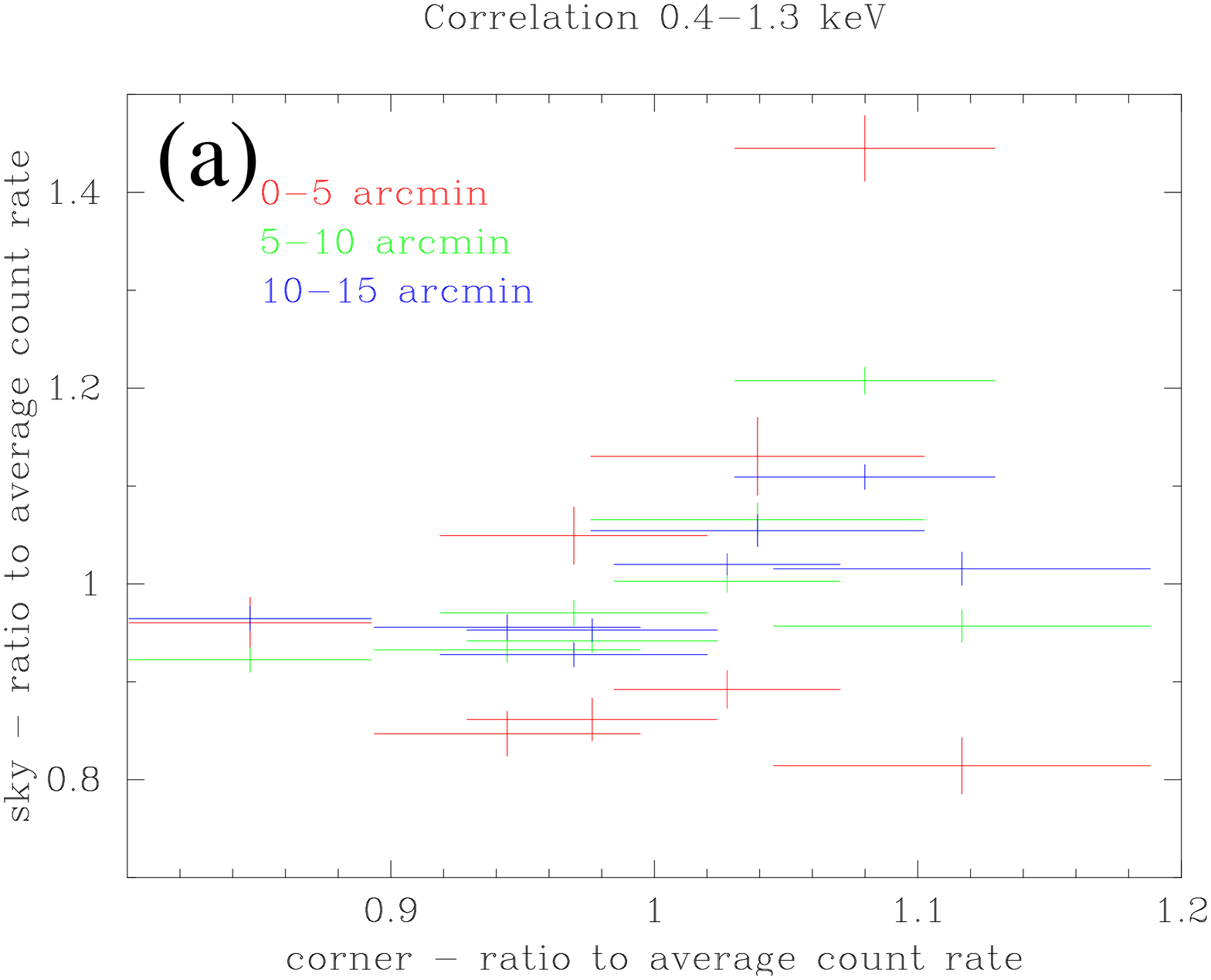}
 \psbox[xsize=0.17#1,ysize=0.17#1]{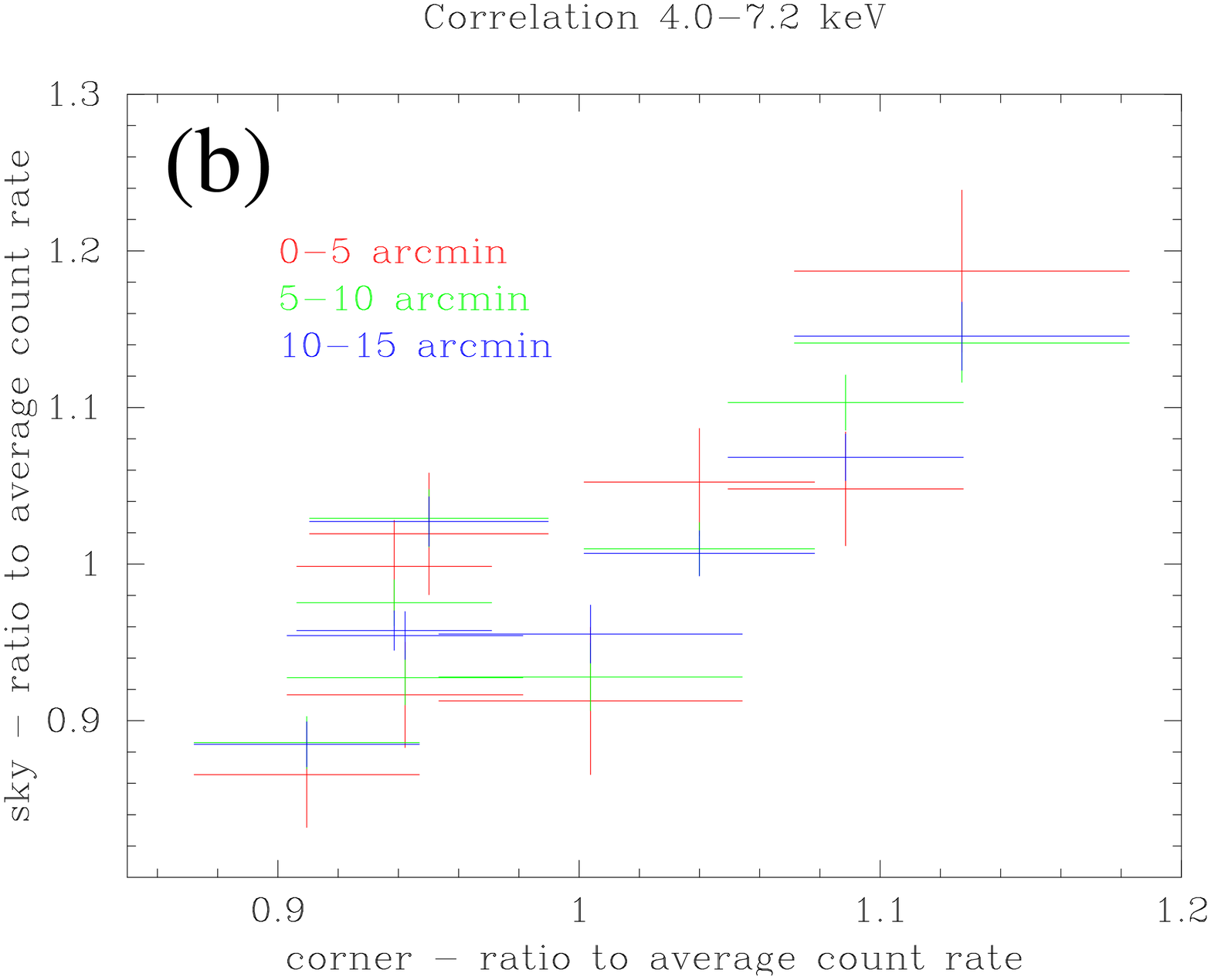}
\vspace*{0.1cm}
 \psbox[xsize=0.17#1,ysize=0.17#1]{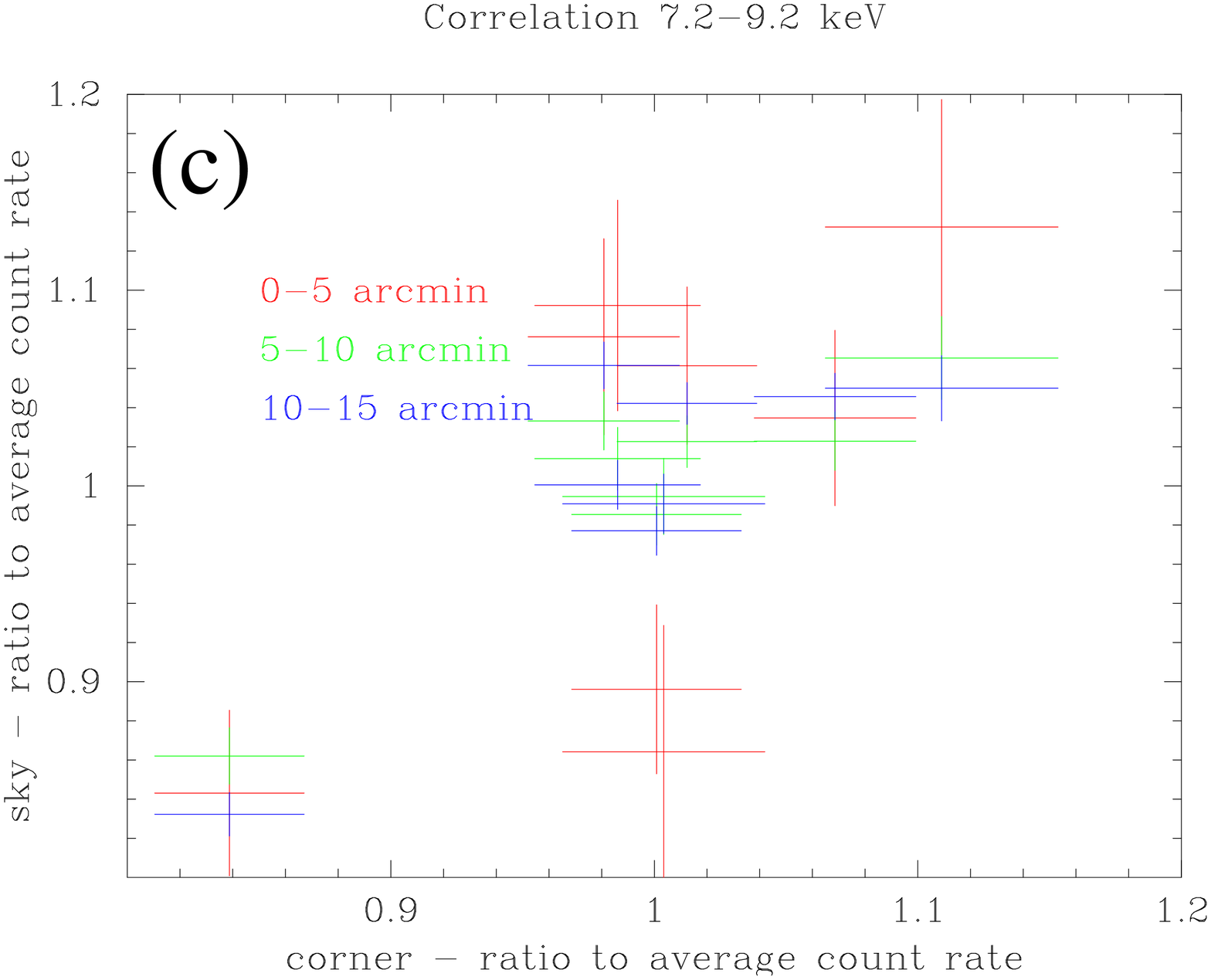}
 \psbox[xsize=0.17#1,ysize=0.17#1]{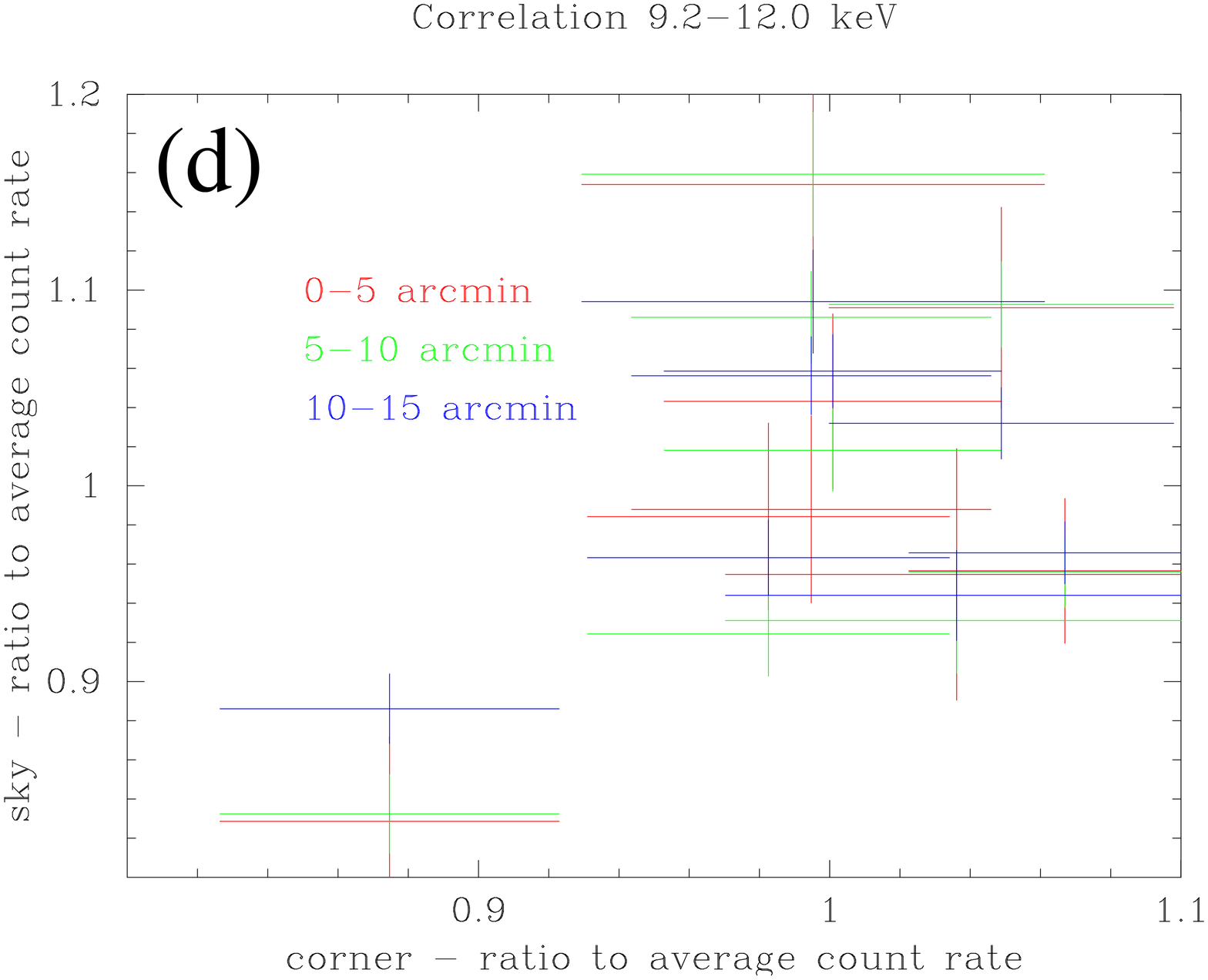}
 \caption{Correlations of PN count rate between the region of outside of FOV
 and within FOV in different energy bands. (a) 0.4--1.3 keV (b) 4.0--7.2 
 keV (c) 7.2-9.2 keV (d) 9.2-12.0 keV.}
 \label{fig:hkatayama-WA3_fig5}
\vspace*{0.5cm}
 \psbox[xsize=0.17#1,ysize=0.17#1]{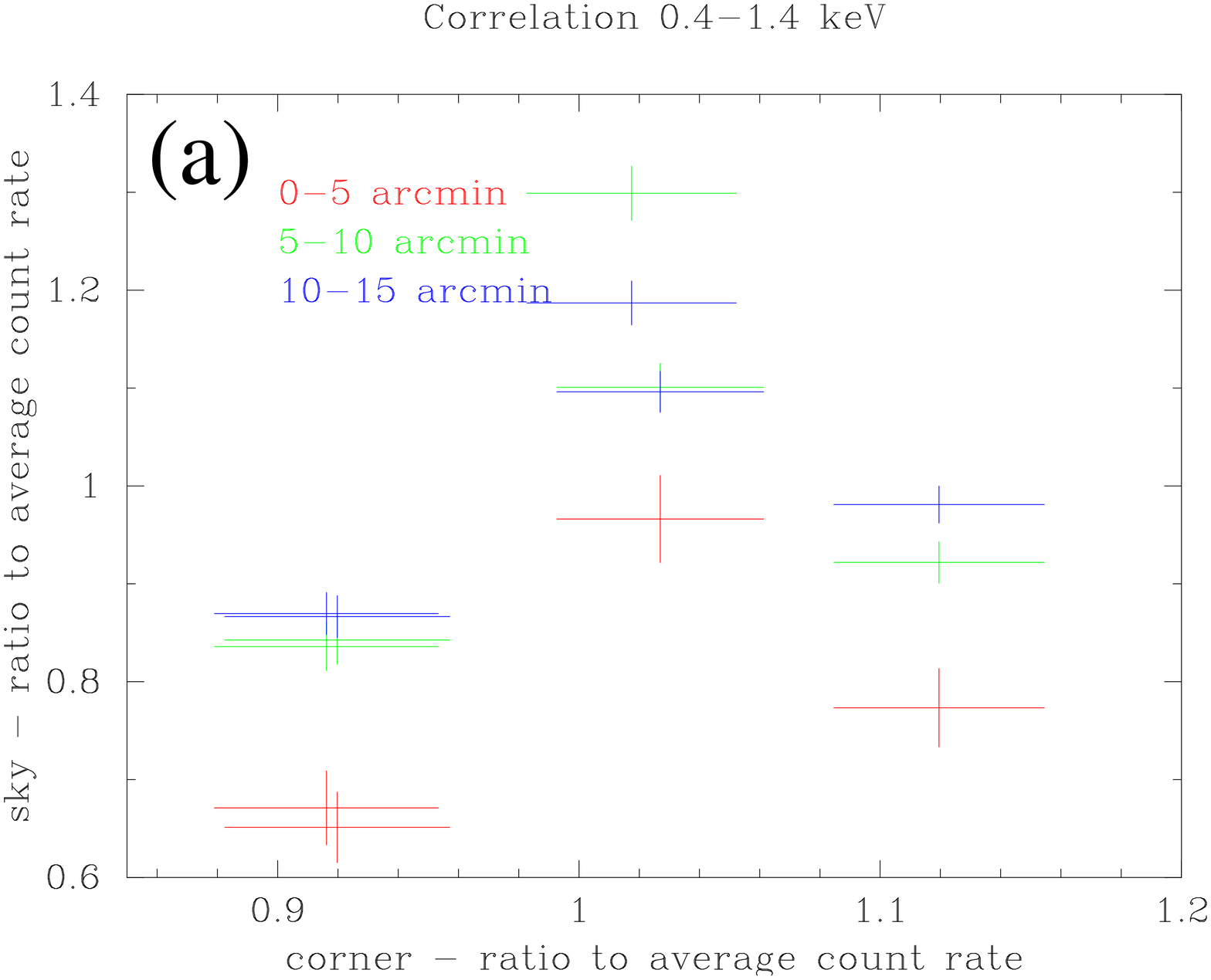}
 \psbox[xsize=0.17#1,ysize=0.17#1]{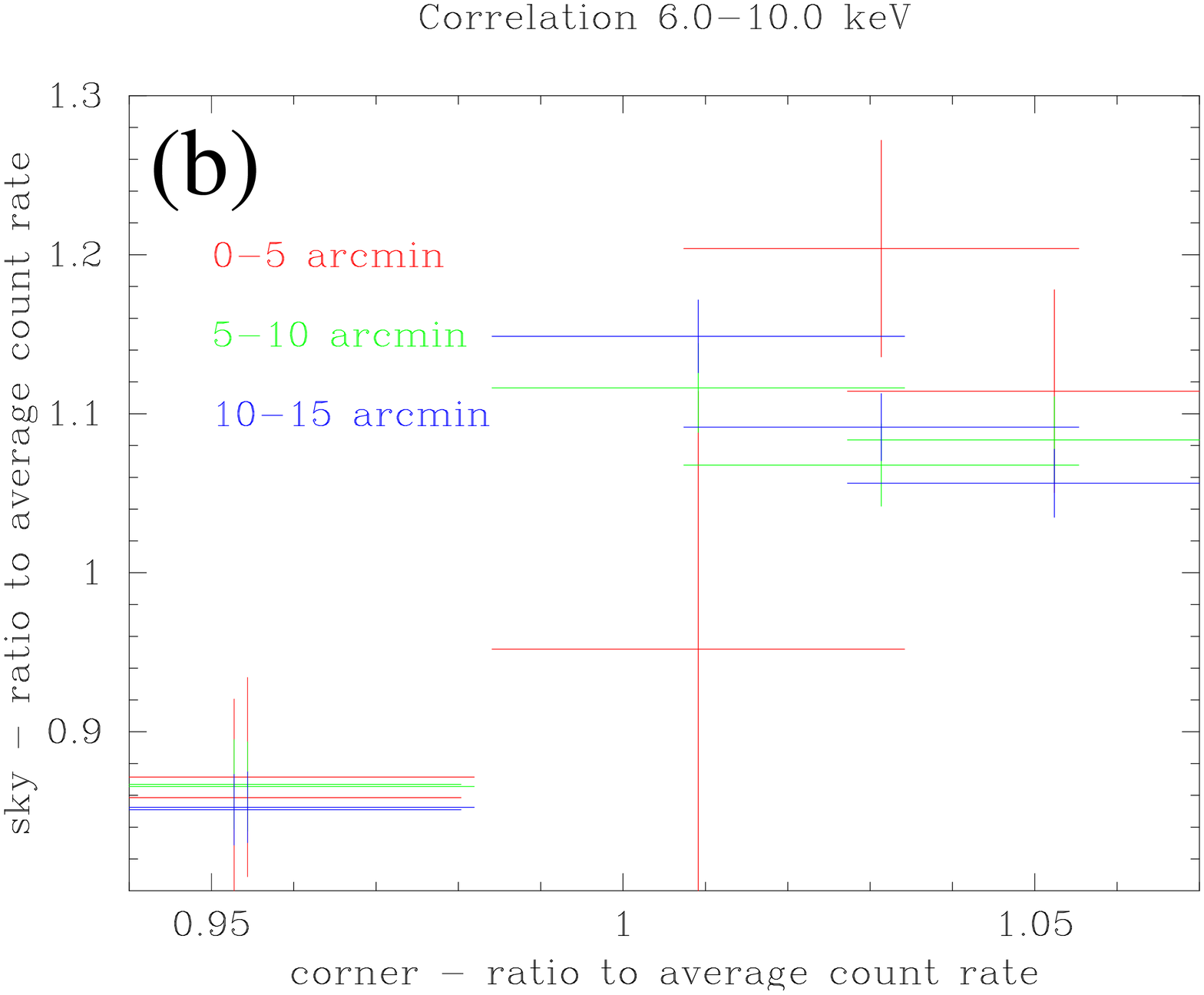}
 \caption{Correlations of MOS count rate between the region of
 outside of FOV and inside FOV in different energy bands. (a) 0.4--1.4
 keV (b) 6.0-10.0 keV.}
 \label{fig:hkatayama-WA3_fig6}
\end{center}
\end{figure}

Figure \ref{fig:hkatayama-WA3_fig5} and figure
\ref{fig:hkatayama-WA3_fig6} display the correlations of count rate
between the region of outside of FOV and inside of FOV in different
energy bands. Red, green, and blue crosses indicate the data sets of
0-5$'$, 5-10$'$, and 10-15$'$, respectively. The PN data set shows a
good correlation in the energy band of 1.7--4.0 keV and 4.0--7.2
keV. Thus, this result represents that the shape of background
distribution does not change in theses energy bands. This also implies
that count rate of outside of FOV is available for background
monitoring. On the other hands, The MOS data set does not show any clear
correlation. As we examined only five data sets for MOS1, further
investigation is needed for MOS.

\section{Long-term variation of background}

\begin{figure}[htbp]
\begin{center}
 \psbox[r,xsize=0.15#1,ysize=0.15#1]{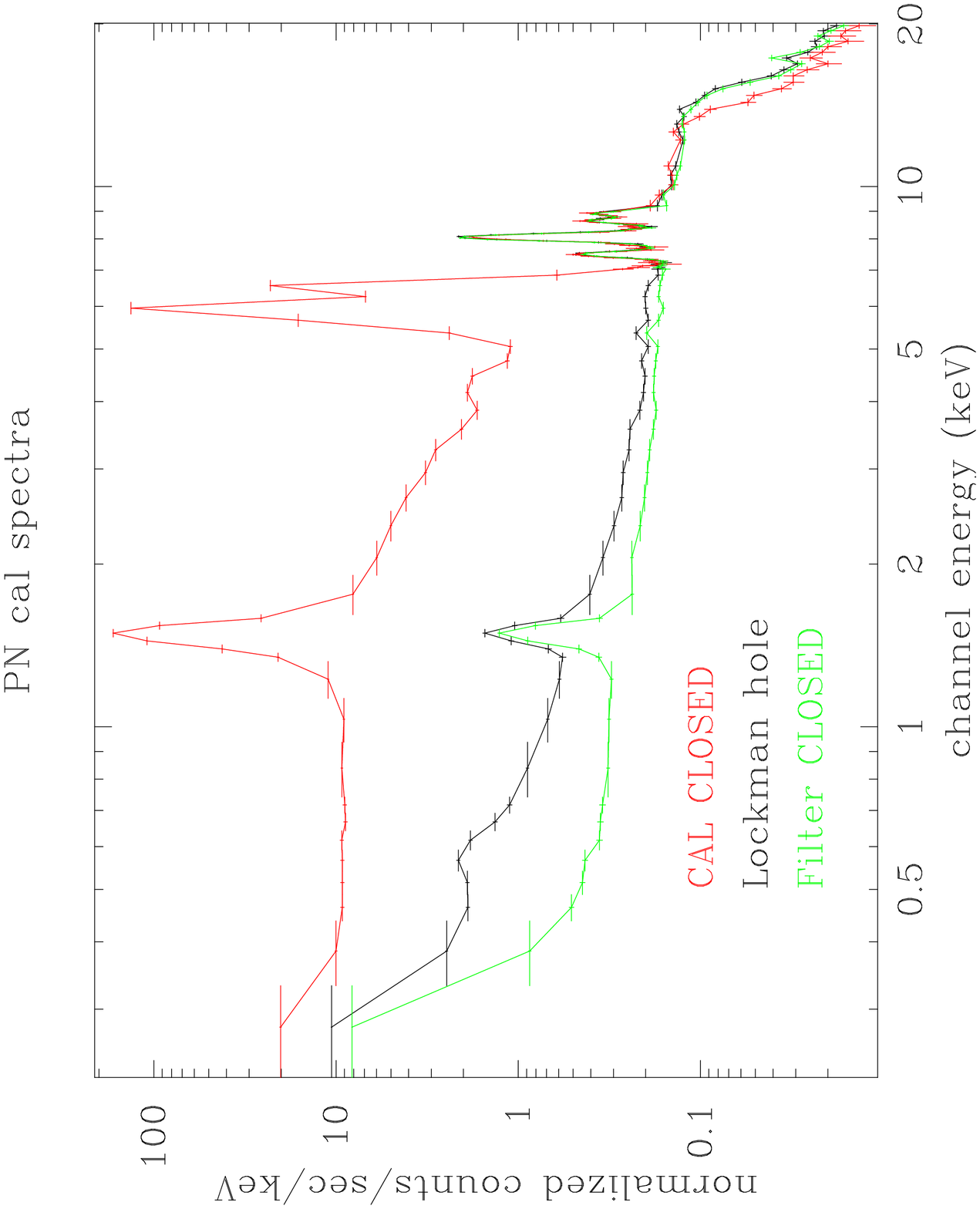}
\hspace*{0.5cm}
 \psbox[xsize=0.20#1,ysize=0.20#1]{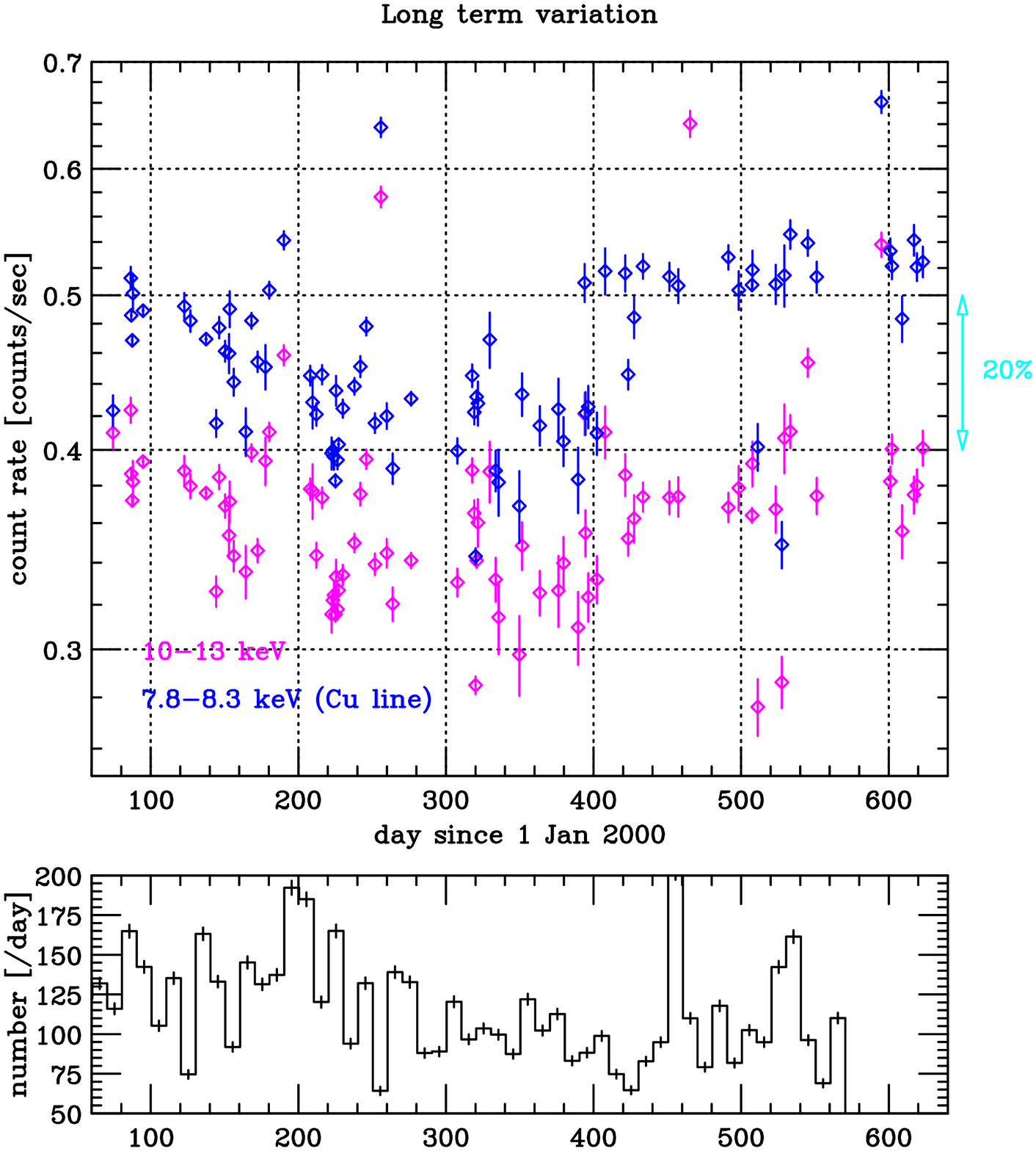}
 \caption{(Right) Spectra of the Lockman hole (Black), filter closed
(Green), and CAL CLOSED data (Red). ~(Left) Long-term variation of the
7.8--8.3 keV and 10.0-13.0 keV count rate. The horizontal axis is the
 day since January 2000. The number of sunspots in the
same term is also displayed in the bottom panel.}
\label{fig:hkatayama-WA3_fig7}
\end{center}
\end{figure}

In order to investigate the long-term variation of the PN background, we
used CAL CLOSED data, which is calibration source data with filter
closed. Figure \ref{fig:hkatayama-WA3_fig7} (Left) displays the spectra
of the Lockman hole, filter closed, and CAL CLOSED data. The 7.0--13.0
keV band is free from events by the calibration source, and its count
rate can be usable for monitoring the background count rate.

The top panel of figure \ref{fig:hkatayama-WA3_fig7} (Right) displays
the long-term light curve in the 7.8--8.3 keV and 10.0--13.0 keV band
derived from the CAL CLOSED data. 
The average background count rate decreased by
20 \% from March 2000 to January 2001, however, it regained in February
2001.  
This long term variation is most probably related to the solar activity.
One could see the similar trend in the variation of the number of sunsplots
shown in the bottom panel of figure \ref{fig:hkatayama-WA3_fig7} (Right).

\section{Parameters for background modeling}

In order to obtain the parameter for background modeling, we investigated
the correlations between characteristic parameters and the 2--10 keV count
rate. 
\begin{enumerate}

 \item[1]{Number of discarded columns}

         For PN, the column that detects an event above upper energy
         threshold and its neighbor columns are removed from the data
         as a background. 
         This is called MIP rejection. As most of these MIP events are
	 cosmic-ray events, the number of columns that are discarded by
	 MIP rejection is expected to be related to the remaining
	 background count rate.

         Figure \ref{fig:hkatayama-WA3_fig8} shows the relation between
	 the 2--10 keV count rates and the number of discarded
	 columns. There is no clear correlation between these two
	 parameters.

	 This might be because the energy spectrum of particles changes
	 temporally. MIP events are attributed to all particles of which
	 energy is above a certain extent. The population of such
	 particles is very large. On the other hands, the 2-10keV
	 background, in particular the continuum, is attributed to a
	 portion of the whole particles that have a certain
	 configuration of the energy and the incident
	 direction. Therefore, the 2-10keV background count rate would
	 fluctuate largely by the temporal change of the particle energy
	 spectrum.

\begin{figure}[htbp]
\begin{center}
 \psbox[r,xsize=0.17#1,ysize=0.17#1]{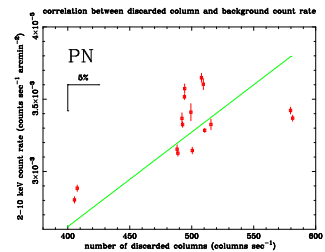}
 \caption{Correlation between the number of discarded columns and the
 2--10 keV count rate.}  \label{fig:hkatayama-WA3_fig8}
\end{center}
\end{figure}

 \item[2]{Count rate of 10--12 keV}

	 For many sources, emissions above 10 keV are negligibly small
	 compared with the background. Thus, the count rate above 10 keV
	 could be usable for the background modeling.

         In figure \ref{fig:hkatayama-WA3_fig9}, the background count
	 rates in 10--12 keV are plotted against those in 2--10 keV. The
	 PN data set shows a good correlation, while the MOS data set
	 does not show a clear correlation.

\begin{figure}[htbp]
\begin{center}
 \psbox[r,xsize=0.17#1,ysize=0.17#1]{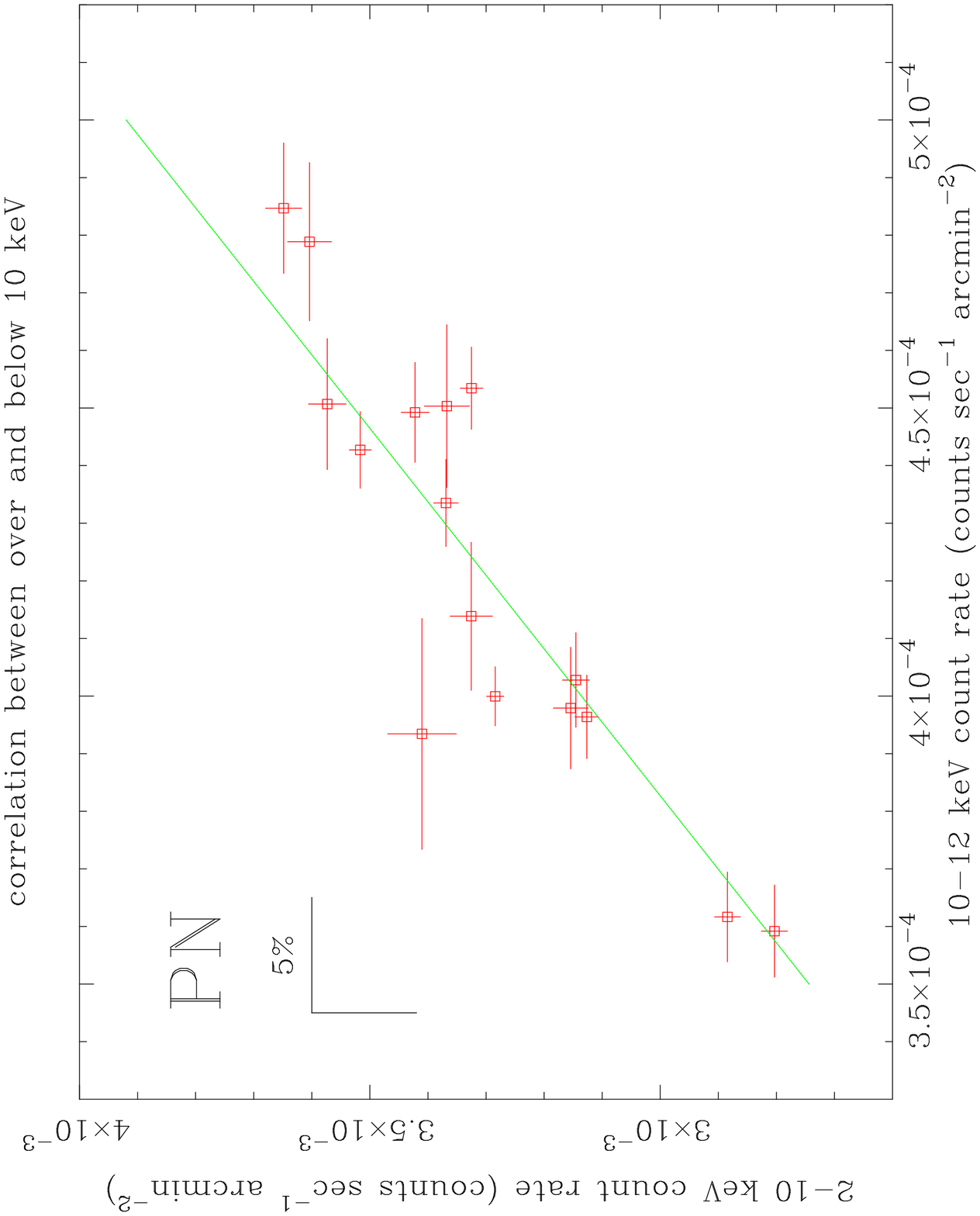}
 \psbox[r,xsize=0.17#1,ysize=0.17#1]{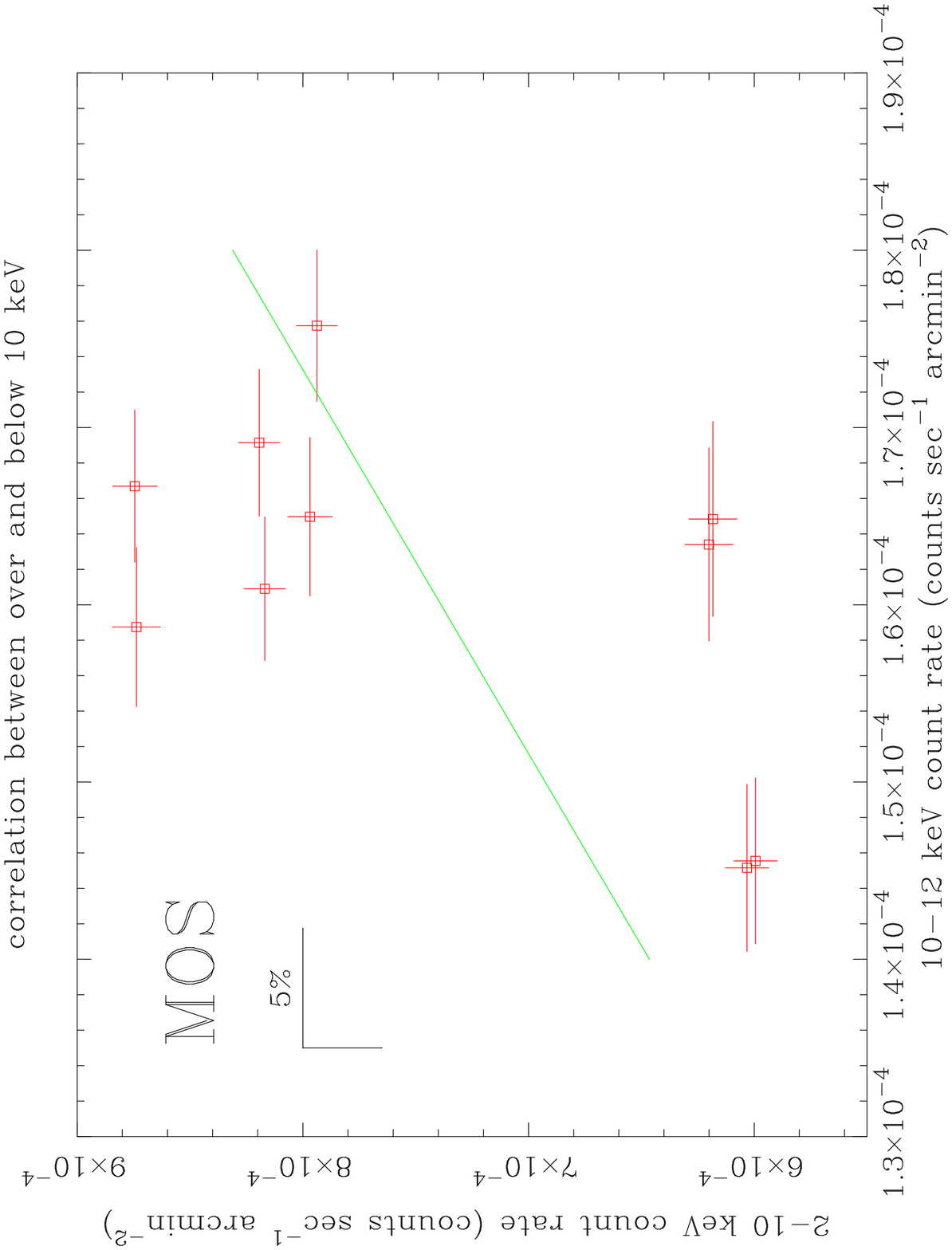}
 \caption{correlations between the 10--12 keV count
	 rate and the 2--10 keV count rate.}
 \label{fig:hkatayama-WA3_fig9}
\end{center}
\end{figure}

 \item[3]{Count rate of outside of FOV}

	 Figure \ref{fig:hkatayama-WA3_fig10} displays the 2--10 keV
	 count rate detected outside of FOV vs. that in FOV. The PN data set
	 shows a good correlation.

	 The good correlation seen for the PN data can be usable for
	 monitoring the background. In order to examine the efficacy,
	 each spectrum shown in figure 3 is renormalized so that the
	 corresponding count rate of outside of FOV becomes the same as
	 that of the average value. Figure \ref{fig:hkatayama-WA3_fig11}
	 shows the fluctuations (standard deviations of 1 $\sigma$) of
	 original spectra and corrected spectra. The variation of PN
	 background spectra decreases from 8 \% to 3 \% in the 2--7 keV
	 band.

\begin{figure}[htbp]
\begin{center}
 \psbox[r,xsize=0.17#1,ysize=0.17#1]{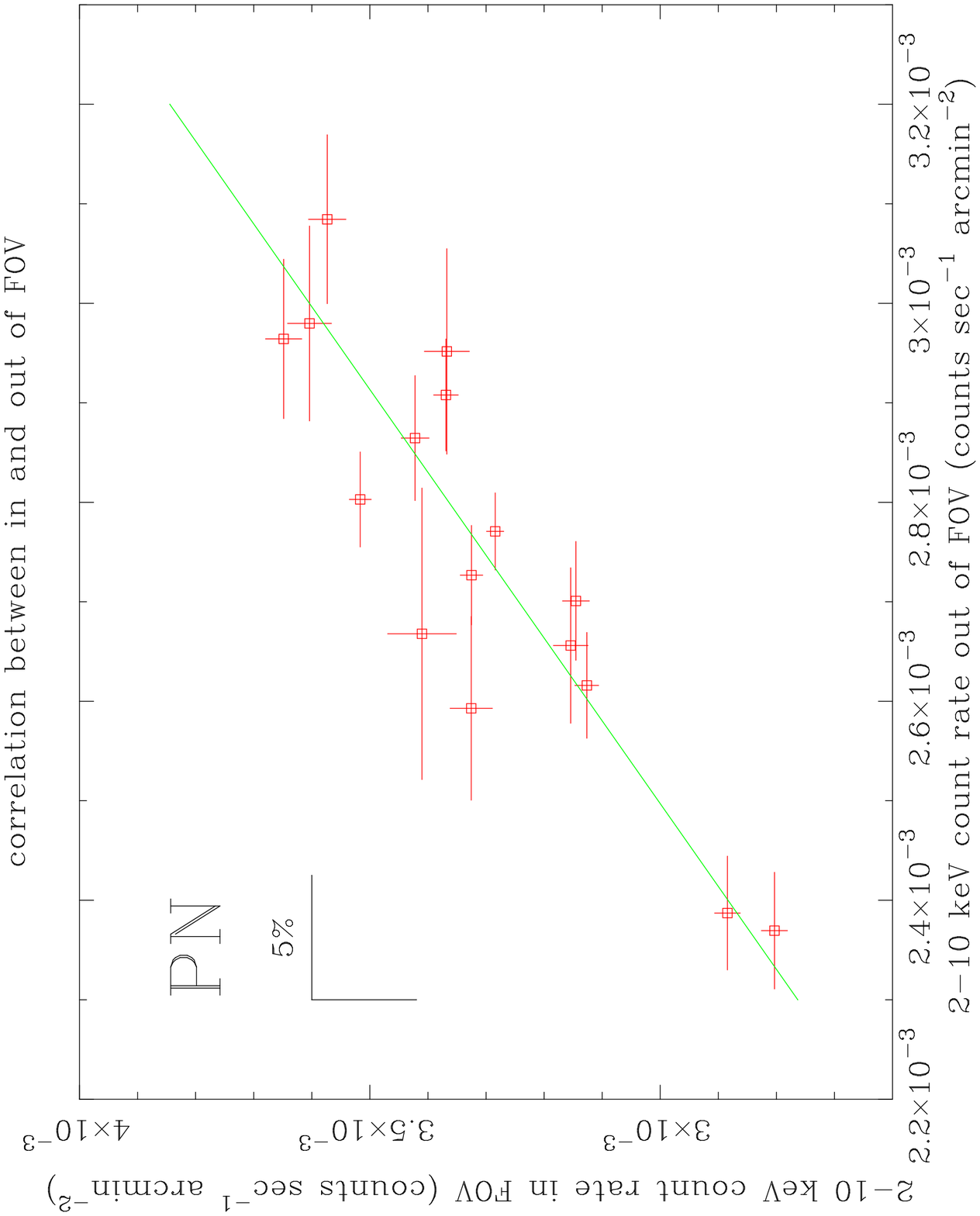}
 \psbox[r,xsize=0.17#1,ysize=0.17#1]{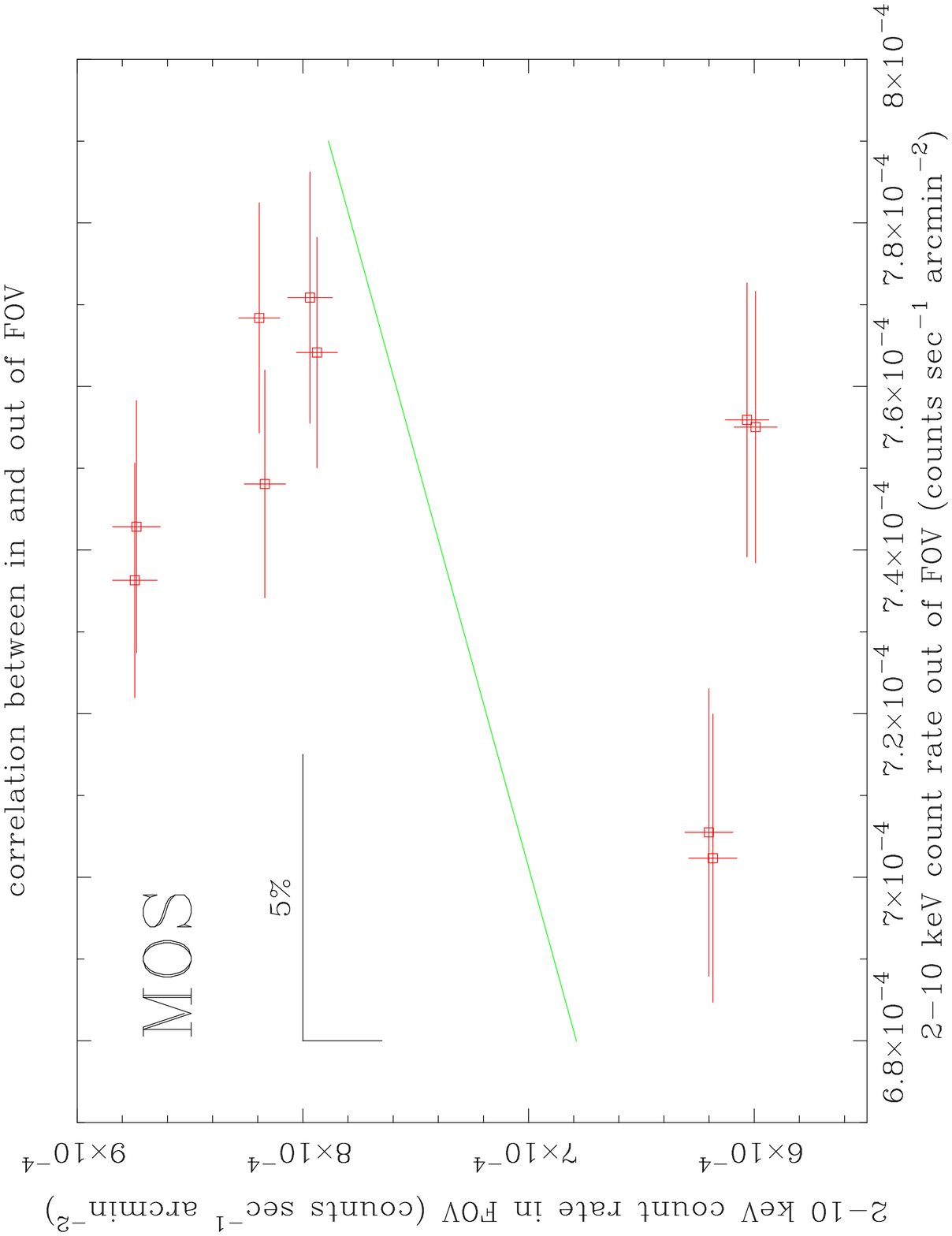}
 \caption{Correlations between the 2--10 keV count rate of outside of FOV
 and that of in FOV.}  
\label{fig:hkatayama-WA3_fig10}
\vspace*{0.5cm}
 \psbox[xsize=0.18#1,ysize=0.18#1]{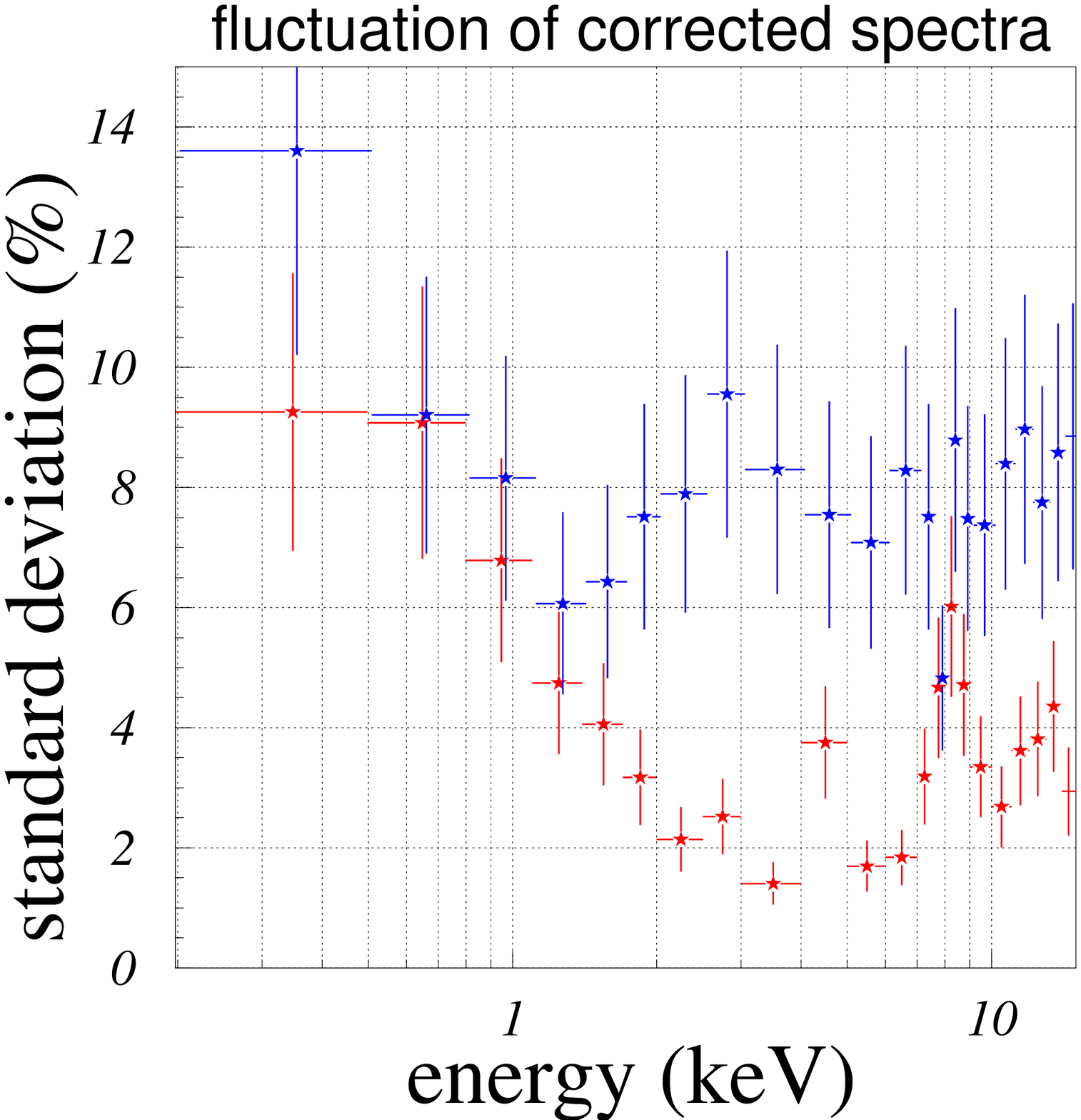}
 \caption{Standard deviations of PN background spectra. Blue data are
 original spectra and Red data are spectra corrected by count rates
 of outside of FOV. The variation decreases from 8 \% to 3 \% in the 2--7
 keV band.}  
\label{fig:hkatayama-WA3_fig11}
\end{center}
\end{figure}

\end{enumerate}

\section{Comparison with ASCA and Chandra}

Finally, we show the comparison of the background spectra among different
X-ray CCDs onboard XMM, ASCA, and Chandra. 
The background spectrum of ASCA SIS is obtained from a blank sky data. 
The Chandra ACIS-S and ACIS-I data used here
are also taken from blank sky observations compiled by
\cite*{hkatayama-WA3:mark01}.

\begin{figure}[htbp]
\begin{center}
 \psbox[r,xsize=0.22#1,ysize=0.22#1]{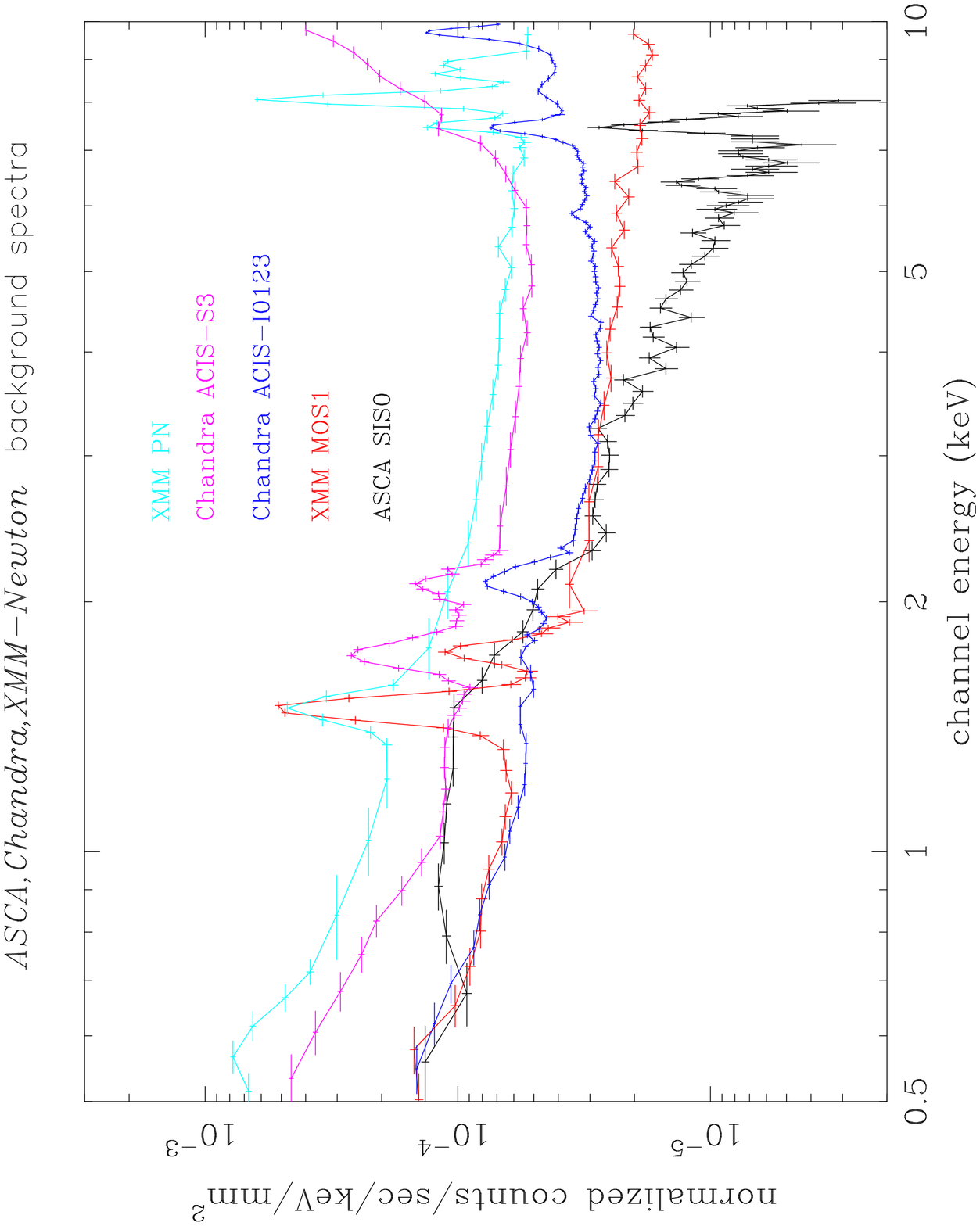}
\vspace*{0.1cm}
 \psbox[r,xsize=0.22#1,ysize=0.22#1]{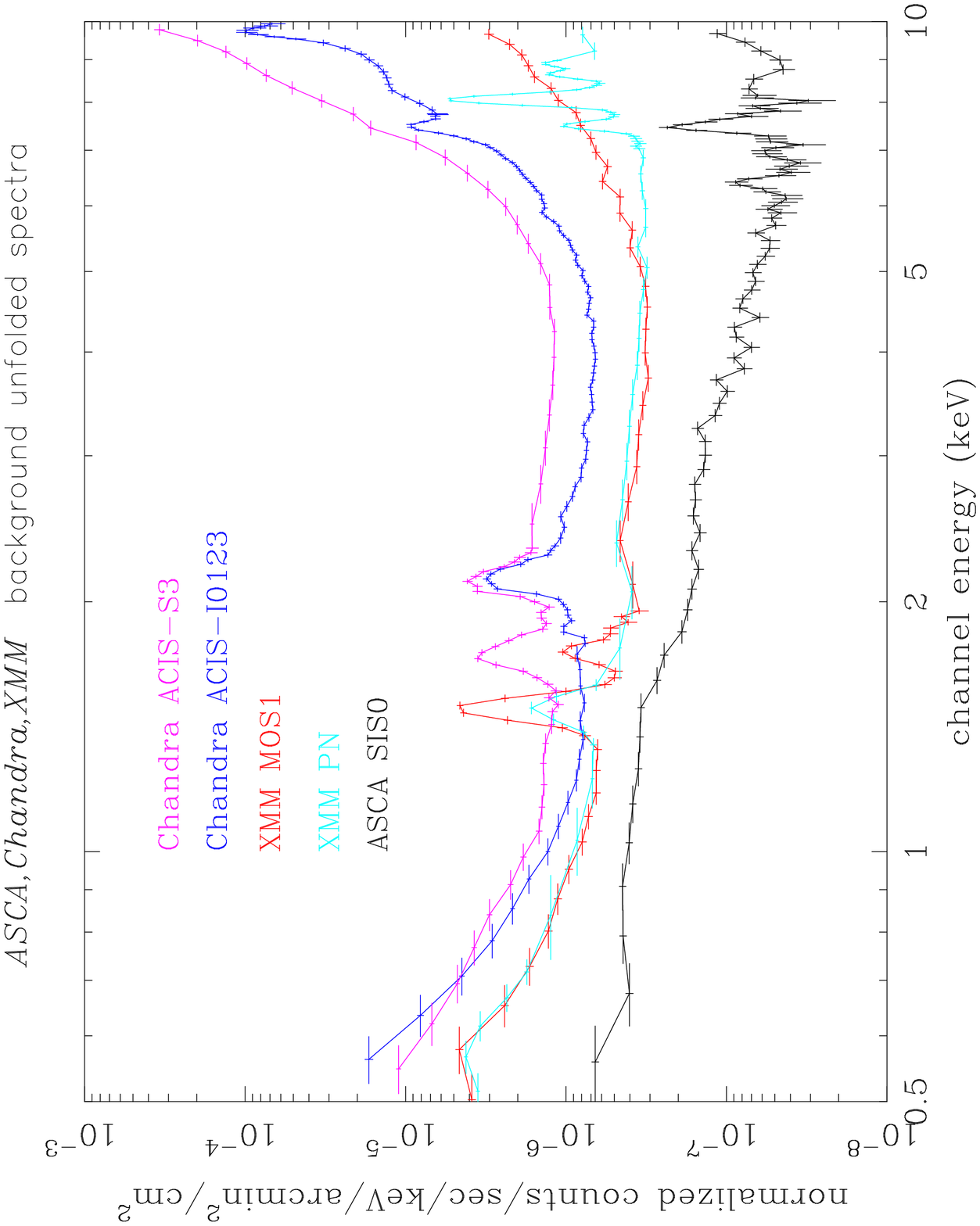}
 \caption{(Top) XMM, Chandra, and ASCA background spectra normalized by
 CCD area. (Bottom) XMM, Chandra, and ASCA background spectra normalized
 by effective area.  } 
\label{fig:hkatayama-WA3_fig12}
\end{center}
\end{figure}

The top panel of figure \ref{fig:hkatayama-WA3_fig12} shows the
background spectra normalized by CCD area, which thus represents the
background count rate per unit area on the focal plane. As these spectra
include CXB component, a direct comparison on the particle event flux is
meaningful only above about 5keV. The difference of the background count
rates among different instruments should be explained by the
cosmic-ray-particle flux depending on the satellite orbit, and the
sensitivity of the CCDs, which is governed by the thickness of the
depletion layer and other structure (e.g. front-illuminated or
back-illuminated).

In the bottom panel of figure \ref{fig:hkatayama-WA3_fig12}, each
spectrum is normalized by the effective area of the X-ray telescope plus
CCD and by the solid angle of the FOV. Thus, it represents a surface
brightness of the background.  ASCA SIS is the most sensitive for faint
diffuse X-ray sources.

\end{document}